\documentclass[aps,prb,reprint,superscriptaddress]{revtex4-1}

\usepackage{amssymb,amsmath,mathptmx}
\usepackage{graphicx}
\usepackage{siunitx}
\usepackage{floatrow}
\usepackage{romannum}
\usepackage{CJK}
\usepackage[breaklinks=true,colorlinks=true,bookmarks=false,urlcolor=blue,
citecolor=blue,linkcolor=blue]{hyperref}
\usepackage{xcolor}

\begin{document}

\title{Andreev reflection in ballistic normal metal/graphene/superconductor junctions}

\author{P. Pandey}
\affiliation{Institute of Nanotechnology, Karlsruhe Institute of Technology, D-76021 Karlsruhe, Germany}

\author{R. Kraft}
\affiliation{Institute of Nanotechnology, Karlsruhe Institute of Technology, D-76021 Karlsruhe, Germany}

\author{R. Krupke}
\affiliation{Institute of Nanotechnology, Karlsruhe Institute of Technology, D-76021 Karlsruhe, Germany}
\affiliation{Department of Materials and Earth Sciences, Technical University Darmstadt, D-64289 Darmstadt, Germany}

\author{D. Beckmann}
\email[detlef.beckmann@kit.edu]{}
\affiliation{Institute of Nanotechnology, Karlsruhe Institute of Technology, D-76021 Karlsruhe, Germany}

\author{R. Danneau}
\email[romain.danneau@kit.edu]{}
\affiliation{Institute of Nanotechnology, Karlsruhe Institute of Technology, D-76021 Karlsruhe, Germany}

\begin{abstract}

We report the study of ballistic transport in normal metal/graphene/superconductor junctions in edge-contact geometry. While in the normal state, we have observed Fabry-P\'{e}rot resonances suggesting that charge carriers travel ballistically, the superconducting state shows that the Andreev reflection at the graphene/superconductor interface is affected by these interferences. Our experimental results in the superconducting state have been analyzed and explained with a modified Octavio-Tinkham-Blonder-Klapwijk model taking into account the magnetic pair-breaking effects and the two different interface transparencies, \textit{i.e.}\,between the normal metal and graphene, and between graphene and the superconductor. We show that the transparency of the normal metal/graphene interface strongly varies with doping at large scale, while it undergoes weaker changes at the graphene/superconductor interface. When a cavity is formed by the charge transfer occurring in the vicinity of the contacts, we see that the transmission probabilities follow the normal state conductance highlighting the interplay between the Andreev processes and the electronic interferometer.

\end{abstract}

\maketitle

\section{Introduction}

In his seminal work, Andreev calculated that an electron has a finite probability to be retro-reflected as a hole at the interface between a normal metal (N) and a superconductor (S) \cite{Andreev1964}. This phenomenon known as Andreev reflection has been extensively studied in normal metal-superconductor junctions \cite{Beenakker1997,Courtois1999,tinkhambook,Klapwijk2004} and strongly depends on the quality of the NS interface \cite{Blonder1982,Octavio1983}. Owing to its electronic band structure and its tunability, graphene in contact with a superconducting material  exhibits gate controlled unusual conductance \cite{Akhmerov2007} and Andreev processes with rich subharmonic gap structures \cite{Cuevas2006} and specular reflection at very low energy \cite{Beenakker2006,Titov2007,Beenakker2008,Komatsu2012,Efetov2016,Efetov2016a}, \textit{i.e.} near the charge neutrality point. This exotic effect could potentially be used to detect valley polarization \cite{Akhmerov2007a} or in spin filters \cite{Greenbaum2007}, while Andreev conversion effects may persist in the quantum Hall regime \cite{Clarke2014,Hou2016,Lee2017,Zhang2019}. Noticeably, an NS interface coupled to a one-dimensional system with strong spin-orbit coupling is expected to host Majorana bound states \cite{Kitaev2001}. 

The development of van der Waals heterostructures designed from two-dimensional materials \cite{Geim2013,Novoselov2016,Yankowitz2019} in addition to the recent progresses in sample fabrication \cite{Dean2010,Wang2013}, dramatically improved both charge carrier mobility and contact transparency in graphene-based electrical devices. As a consequence, the study of proximity induced superconductivity regained interest with the possibilities to measure large supercurrents and ballistic interferences \cite{Calado2015,BenShalom2016,Amet2016,Borzenets2016,Zhu2016,Allen2017,Nanda2017,Zhu2018,Park2018,Kraft2018,Schmidt2018,Kroll2018,Draelos2019,Wang2019}. However, most of the studies related to Andreev processes in this system usually consist of probing the dissipationless current and multiple Andreev reflection in graphene connected to two superconducting contacts, while Andreev bound states in graphene have been detected by tunnelling spectroscopy experiments \cite{Bretheau2017}. Indeed, very few utilize devices with a single superconducting-graphene interface \cite{Efetov2016,Popinciuc2012,Han2018} that can be described by the Blonder-Tinkham-Klapwijk (BTK) model \cite{Blonder1982}. To our knowledge, there is no experimental study on normal metal-graphene-superconductor (NGS) junctions in the ballistic regime.  

Here we investigate NGS devices based on graphene-hexagonal boron nitride (hBN) van der Waals heterostructures. We observe that while the normal state conductance exhibits Fabry-P\'{e}rot (FP)-like interference, the electronic transport and the Andreev processes at the GS interface in the superconducting state are very well described by our modified Octavio-Tinkham-Blonder-Klapwijk (OTBK) model \cite{Blonder1982,Octavio1983,Golubov1995,Perez-Willard2004}. We show that the transmission probabilities extracted from the experimental data follow the same oscillating trend as the normal state conductance which reflects the ballistic nature of the electronic transport in these clean systems. 

\section{Sample design, fabrication process and experimental details}

\begin{figure}
\includegraphics[width=0.9\textwidth]{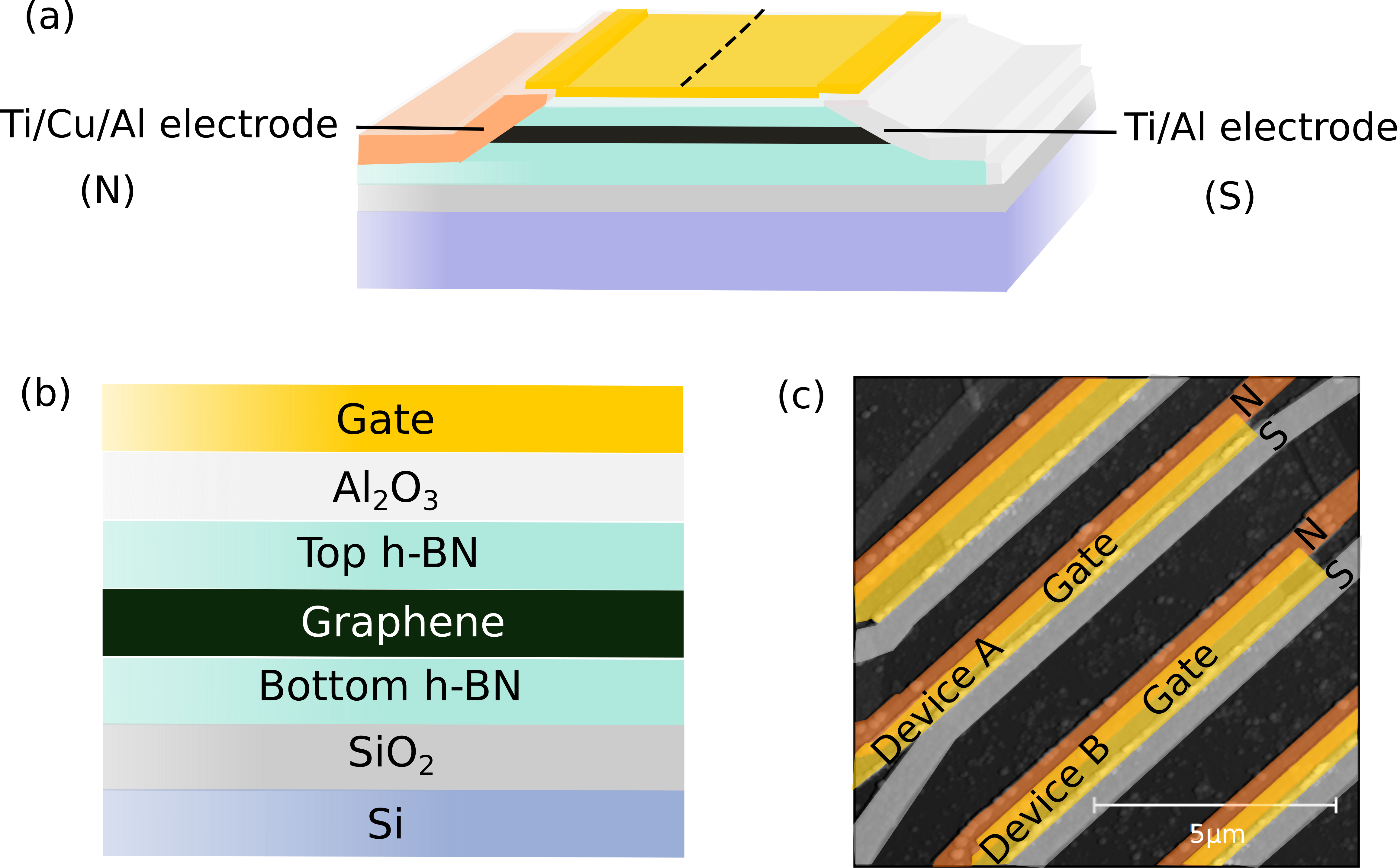}
\caption{(Color online) (a) Schematic of the device geometry. (b) Cross section schematic of the studied devices across the dashed line in (a). (c) False color atomic force micrograph of the two devices with normal (N) and superconducting (S) contacts. Scale bar is 5\,$\mu$m.}
\label{fig:devicefab}
\end{figure}

We have fabricated our van der Waals heterostructure using a dry transfer method following a similar technique as described in Wang \textit{et al.} \cite{Wang2013}. Graphene and hBN crystallites (16 and 20\,nm thick for the top and bottom hBN layer, respectively) were obtained by exfoliation of natural graphite (from NGS Naturgraphit GmbH) and commercial hBN powder (Momentive, grade PT110), respectively and transferred on Si substrates with 300~nm thick SiO$_2$ top layer. Edge contacts to the graphene sheet were established in a self-aligned manner by etching the desired connection area followed by metallization as described in Kraft \textit{et al.} \cite{Kraft2018}. However, this process had to be adapted in order to deposit two different metals. The contact geometry is defined by standard e-beam lithography where poly-methyl methacrylate (PMMA) was used as a mask, followed by reactive ion etching (RIE) with a plasma of CHF$_3$ and O$_2$ through the van der Waals heterostructure to define the first edge contact to the graphene sheet. Immediately after RIE, the first electrode metal is evaporated under ultra-high vacuum (pressure of the order of 10$^{-9}\,$mbar). This procedure is repeated as two different metals have to be deposited for the contact electrodes. We used Ti/Al (5 and 70~nm, respectively) for the superconducting electrode and Ti/Cu/Al (5, 70 and 5~nm, respectively) for the normal metal electrode. The Al and Cu contacts are parallel. While Ti serves as an adhesive layer, the 5\,nm Al in the normal electrode serves as a capping layer to protect Cu from oxidation. In the final fabrication step, a 25~nm thick Al$_2$O$_3$ layer was deposited on top of these devices by atomic layer deposition, followed by the fabrication of a top gate made of Ti/Cu/Al (5, 80 and 5~nm, respectively).

Schematics of the devices are shown in Fig.~\ref{fig:devicefab}(a) and (b) and an AFM micrograph of the measured devices (A and B) is shown in Fig.~\ref{fig:devicefab}(c). These devices have high width over length $W/L$ ratio with dimensions of W = 6\,$\mu$m, L = 0.25\,$\mu$m and W = 5\,$\mu$m, L = 0.45\,$\mu$m, respectively. The data presented in the main text is from device A unless stated otherwise. Similar data was obtained from device B except the FP resonances which were observed only in device A. Electrical characterization of these devices was carried out in a $^3$He/$^4$He dilution refrigerator at a base temperature $T$ in the range of 20-100~mK unless otherwise mentioned. All of the measurements were conducted in a pseudo four-probe configuration with standard low frequency lock-in detection technique. 

\section{Ballistic transport with asymmetrical contact in the normal state}

\begin{figure}
\includegraphics[width=0.8\textwidth]{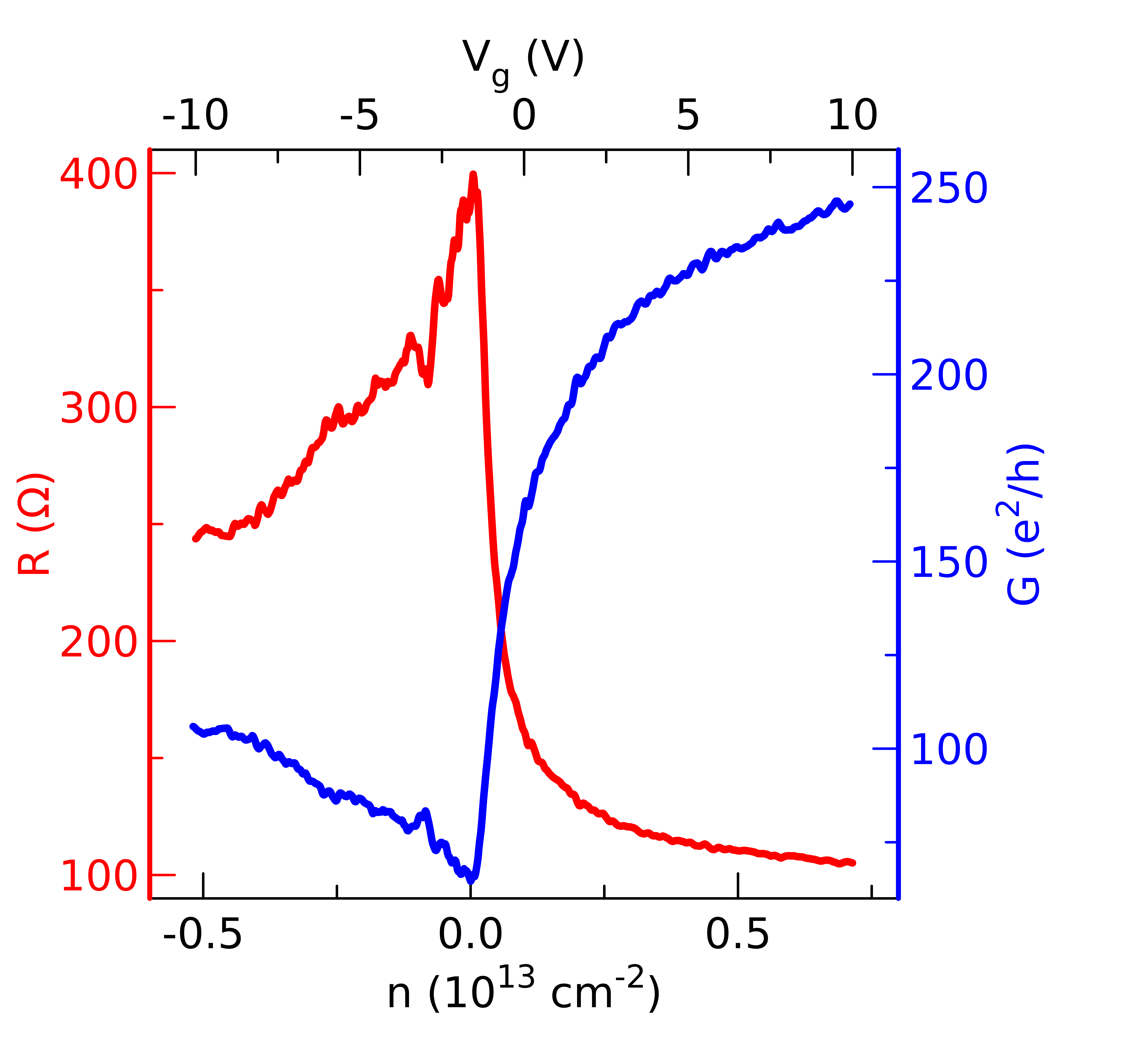}
\caption{(Color online) Resistance $R$ and conductance $G$ versus charge carrier density (n) and gate voltage ($V_\mathrm{g}$).}
\label{fig:Resistance}
\end{figure}

In this section, we show how the asymmetry of the contacts tunes the FP resonances characterizing the ballistic regime of our two-terminal devices in the normal state.  Fig.~\ref{fig:Resistance} displays both resistance $R$ and conductance $G$ of device A as a function of gate voltage $V_\mathrm{g}$ and charge carrier density $n$ at $T \sim 4.2$\,K. The charge carrier density $n$ was obtained from Shubnikov-de Haas measurements as a function of gate voltage $V_\mathrm{g}$. We observe asymmetry in the curves \textit{i.e.} higher resistance in the hole-doped (p) region than the electron-doped (n) region, and shift of the charge neutrality point (CNP) towards negative gate voltage. It indicates n-type doping of the graphene sheet and formation of a potential barrier at the contact while driving the Fermi level in the valence band \cite{Giovannetti2008,Khomyakov2009,Khomyakov2010}. We note that as the two contacts are made of different materials (Ti/Cu/Al and Ti/Al), the charge transfer induced by the different work function of the leads might enhance this asymmetry. Since the graphene sheet is n-type doped by the metal contacts, driving the charge transport in graphene to hole-doped regime using the gate results in the formation of pn junctions in the vicinity of the metal electrodes. These two pn-junctions act as partially transmitting interfaces similar to the mirrors in a FP interferometer. As a result, charge carriers are reflected back and forth at the graphene/metal interfaces. Quantum interference takes place due to multiply reflected charge carrier trajectories. Varying the gate voltage changes the Fermi wavelength which leads to an alternating constructive and destructive interference pattern in conductance. For a ballistic device, it can be observed as a periodic oscillation of conductance/resistance while tuning the Fermi wavelength. These oscillations can be clearly seen in the p-doped region in Fig.~\ref{fig:Resistance} and can be attributed to FP interference as a cavity is formed while the Fermi level is positioned in the valence band \cite{katsnelsonbook,Miao2007,Shytov2008,Young2009,Cho2011,Grushina2013,Rickhaus2013,Varlet2014,Du2018}. 

\begin{figure*}
\includegraphics[width=0.9\textwidth]{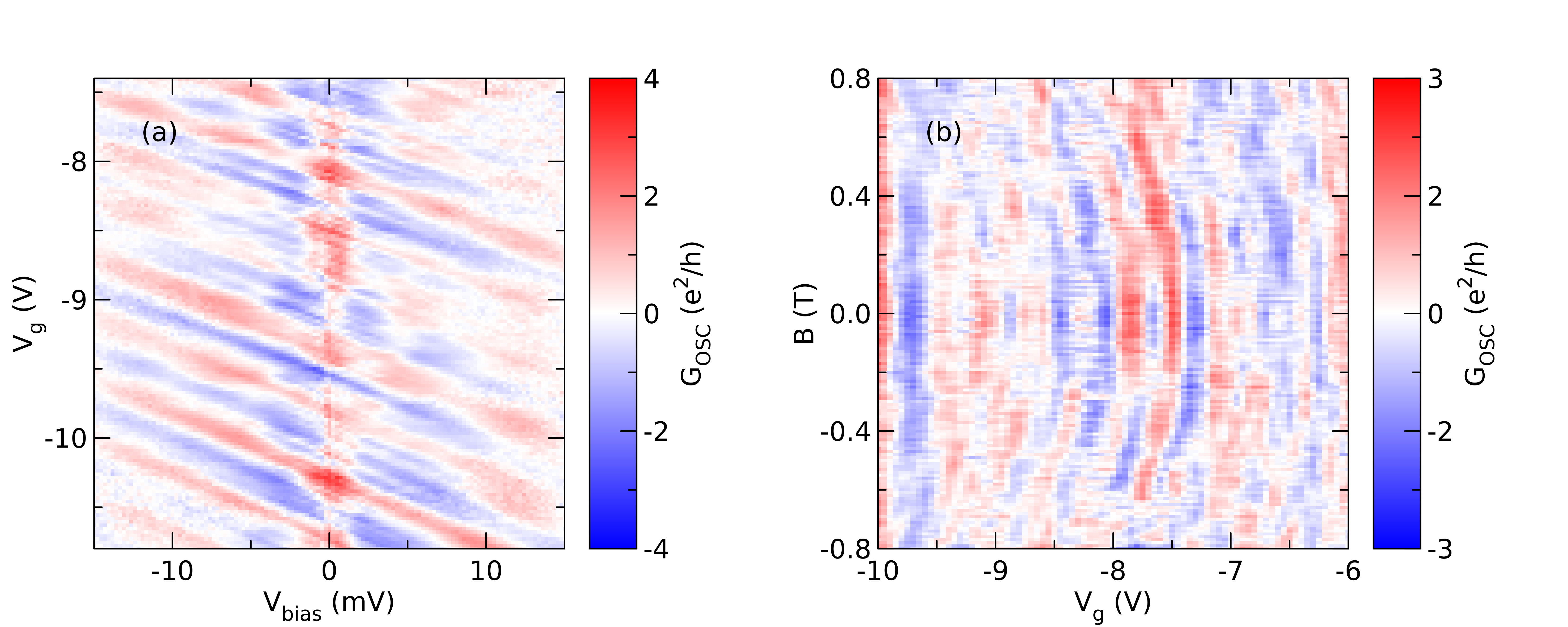}
\caption{(Color online) Energy dependence and magnetic field dispersion of the FP interference in the hole-doped region: (a) $G_{\mathrm{osc}}$ versus applied bias voltage $V_{\mathrm{bias}}$ and gate voltage $V_\mathrm{g}$ showing asymmetric pattern of the FP resonances at temperature $T=100$~mK and magnetic field $B=50$~mT. (b) $G_{\mathrm{osc}}$ versus magnetic field $B$ and gate voltage $V_\mathrm{g}$.}
\label{fig:FigFP}
\end{figure*}

In the simplest case, for devices with high aspect ratio, FP resonances occur when the condition $k_\mathrm{F}L_\mathrm{C} = n\pi$ is satisfied, where $k_\mathrm{F}$ is the Fermi wavevector, $L_\mathrm{C}$ is the cavity length and n is an integer. Considering normal incidence of charge carriers at a fixed bias, $L_{C}$ can be estimated by the change in the Fermi wavevector/wavelength between two consecutive conductance maxima/minima. We have used both bias spectroscopy and magnetic field dependence to characterize the FP interference. As the conductance profile strongly varies when the gate voltage $V_\mathrm{g}$, bias voltage $V_\mathrm{bias}$ or magnetic field $B$ are tuned \cite{Young2009,Du2018}, we have extracted the oscillatory part of the conductance $G_{\mathrm{osc}}$ by subtracting a non-oscillatory background contribution in order to study this interference effect. Fig.~\ref{fig:FigFP}(a) shows the bias spectroscopy map of $G_{\mathrm{osc}}$ versus gate and bias voltages, $V_\mathrm{g}$ and $V_{\mathrm{bias}}$, respectively. We note that the usual checkerboard pattern observed in bias spectroscopy experiments \cite{Miao2007,Cho2011,Grushina2013} is very asymmetric here as it was observed in carbon nanotubes with energy-dependent transmission coefficients implying strong asymmetry between the leads \cite{Wu2007}. Here, the two different materials used to contact the graphene can possibly explain our bias spectroscopy pattern as the two interfaces most likely have different transparencies. We note that the asymmetry of the contacts might affect the visibility of these quantum oscillations. The interference pattern observed here corresponds to a cavity length $L_\mathrm{C}$ of ~240 $\pm$ 5 nm which is consistent with the geometrical dimension of the device, confirming that the cavity is formed by the pn-junctions arising from the charge transfer in the vicinity of the contacts \cite{Giovannetti2008,Khomyakov2009,Khomyakov2010}.  

Thanks to the angle-dependence of the transmission through a potential barier in graphene, an effect known as Klein tunnelling \cite{katsnelsonbook}, FP interference can also be tuned and studied by applying a perpendicular magnetic field $B$. Fig.~\ref{fig:FigFP}(b) displays the effect of low magnetic field on the FP interference pattern before entering the quantum Hall regime, where a parabolic dispersion of the interference can be observed due to the Aharonov-Bohm phase.  As already observed in FP interferometers \cite{Young2009,Grushina2013,Rickhaus2013}, when $B$ increases, the charge carrier trajectories within the cavity bend up to the point that, in the momentum space, the path encloses the origin picking up a non-trivial Berry phase of $\pi$ \cite{Young2009}. Experimentally, this appears as a phase shift of $\pi$ in the conductance/resistance oscillations \cite{Young2009,Grushina2013,Rickhaus2013} at a certain magnetic field as seen in Fig.~\ref{fig:FigFP}(b). It is to be noted that this measurement was carried out in another cool-down which accounts for the slight change in the position of conductance maxima/minima as compared to Fig.~\ref{fig:FigFP}(a) with respect to the applied gate voltage.

\section{Andreev reflection and interface analysis in NGS junctions}

In the previous section we have seen that the FP interference in the normal state is affected by the asymmetry of the leads. Here, we investigate the superconducting state and observe that the ballistic nature of the quasiparticle transport is also visible at the graphene/superconductor interface. In order to study superconductivity in these NGS junctions, we measured series of differential conductance $dI/dV$ spectra as a function of gate voltage $V_\mathrm{g}$, perpendicular magnetic field $B$ and temperature $T$, and compare our data with a modified OTBK model \cite{Octavio1983} combined with Ref.~\onlinecite{Perez-Willard2004} (without accounting for different spin density of states). We then analyze the resulting fitting parameters and observe that the interface transparencies follow the oscillatory behavior of the conductance due to FP interference as an asymmetrical cavity is built in our devices.  

\subsection{Modeling}

\begin{figure}[htb]
\includegraphics[width=1\textwidth]{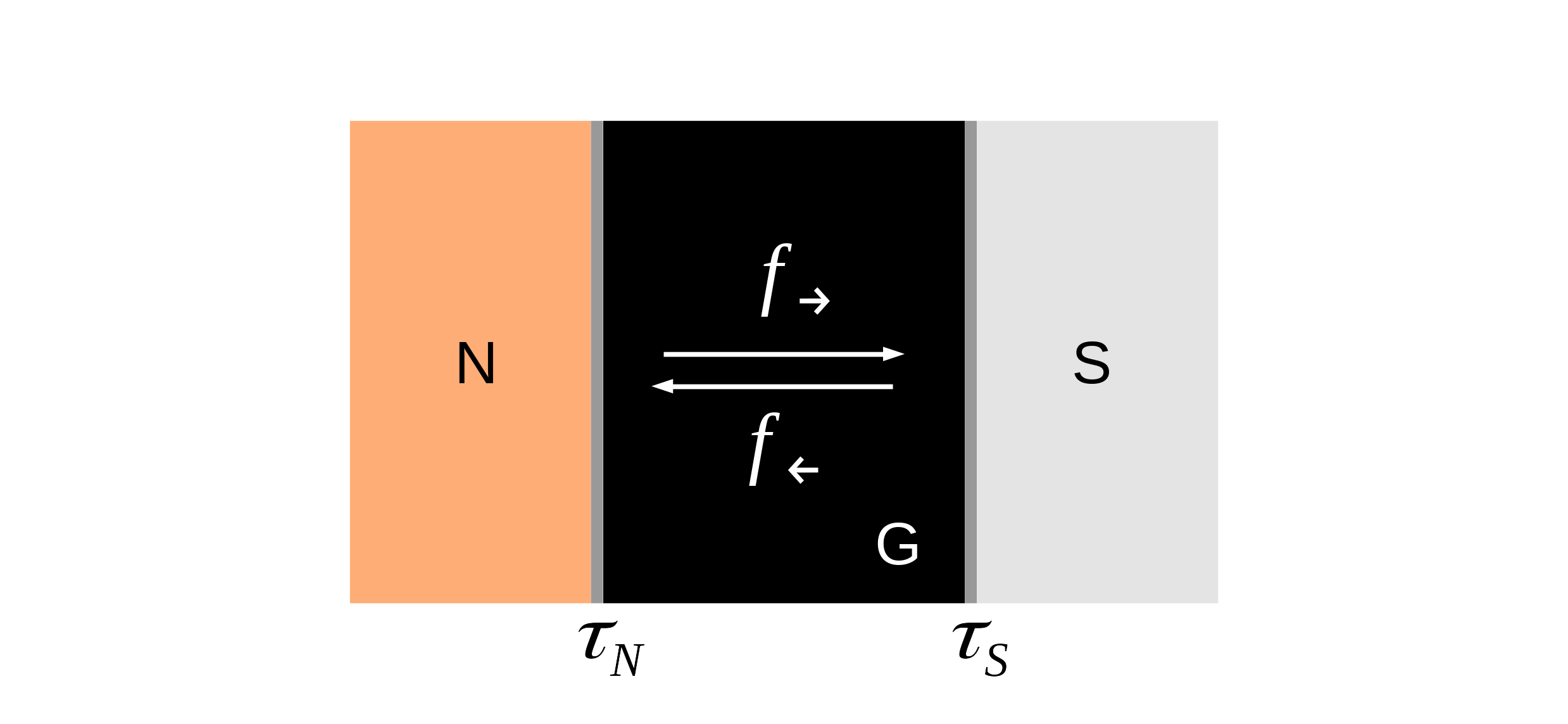}
\caption{(Color online) Schematic view of the model for an NGS junction (see text).}
\end{figure}

\begin{figure*}[htbp]
\includegraphics[width=1\textwidth]{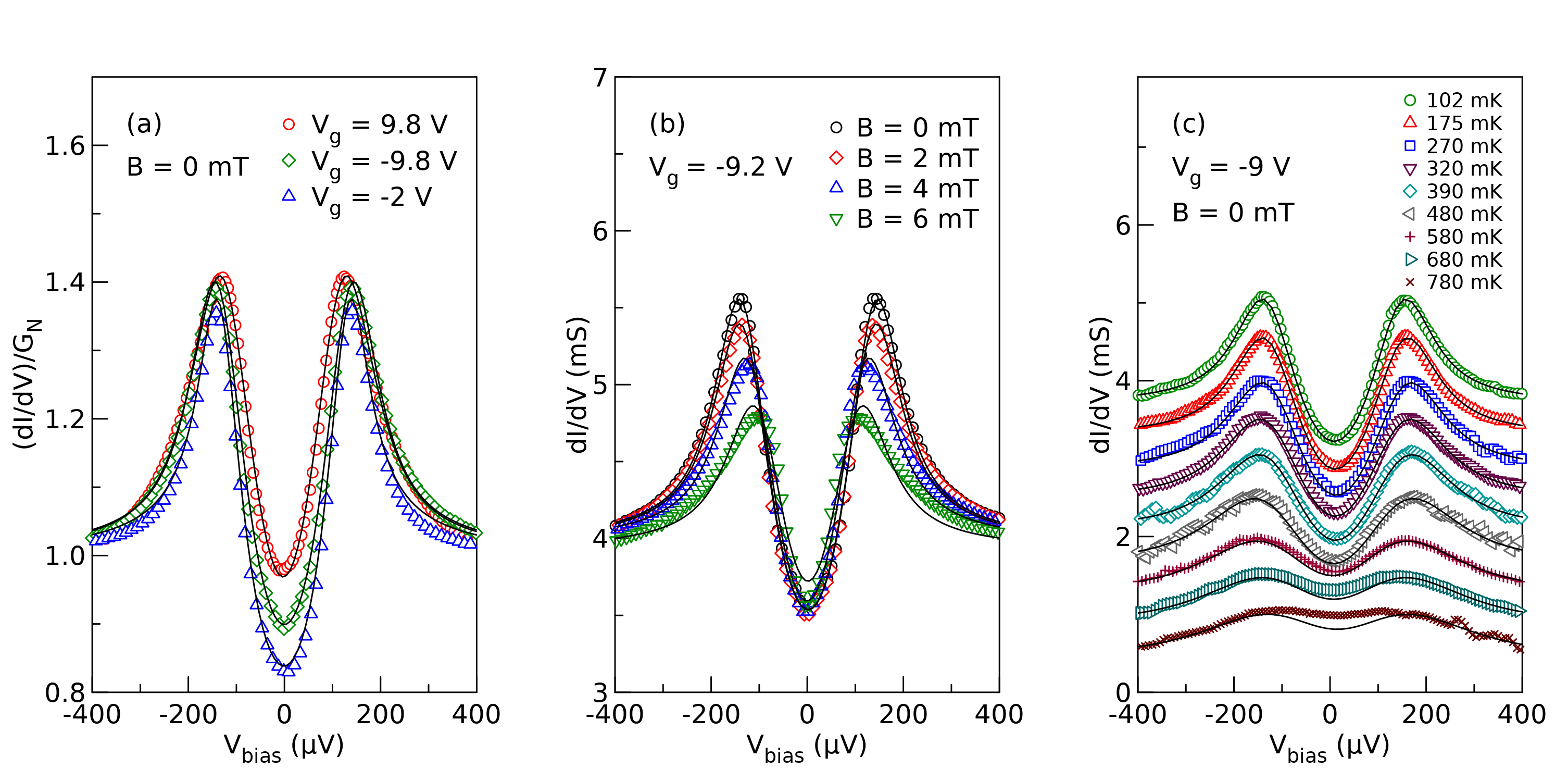}
\caption{(Color online) Differential conductance $dI/dV$ measured as a function of the applied bias voltage $V_{\mathrm{bias}}$ for: (a) normalized by the normal state conductance $G_\mathrm{N}$ (from the measurement at 4.2\,K) for applied gate voltage $V_\mathrm{g}$ at $T \sim 20$\,mK corresponding to electron-doped region (circle), hole-doped region (diamond) and close to the charge neutrality point (triangle). (b) under various values of perpendicular magnetic field $B$ at $T \sim 20$\,mK and $V_\mathrm{g} =-9.2$\,V corresponding to the hole-doped region. (c) for different temperatures at $V_\mathrm{g} =-9$\,V, \textit{i.e.}\,in the the hole-doped region. In all panels, the symbols represent experimental data and the solid curves represent the best fit of the data with our model.} 
\label{fig:superconducting}
\end{figure*}

To account for the two separate interfaces of our NGS structure, we model the system in the superconducting state using a modified OTBK model \cite{Octavio1983}. The graphene sheet (G) is modeled as an ideal conductor, connecting a superconducting (S) terminal (right) and a normal metal (N) terminal (left). The interfaces have normal-state transmission probabilities $\tau_{S}$ and $\tau_{N}$, respectively. These transmission probabilities summarily account for the material interfaces and the pn junctions formed in the p-doped region. The superconducting terminal is at electrochemical potential $0$, and the normal metal terminal is at electrochemical potential $\mu=-eV$. To calculate the current through the device, we determine the left- and right-moving distribution functions $f_\leftarrow(\epsilon)$ and $f_\rightarrow(\epsilon)$ from the boundary conditions
\begin{eqnarray}
 f_\leftarrow(\epsilon) & = & T(\epsilon)f_0(\epsilon) + R(\epsilon)f_\rightarrow(\epsilon) + A(\epsilon)\left(1-f_\rightarrow(-\epsilon)\right), \nonumber \\
 f_\rightarrow(\epsilon)  & = &\tau_{N} f_0(\epsilon-\mu) + r_{N} f_\leftarrow(\epsilon), \nonumber
\end{eqnarray}
where $T(\epsilon)$, $R(\epsilon)$ and $A(\epsilon)$ are the probabilities for normal transmission, normal reflection and Andreev reflection, respectively, at the superconducting terminal, $r_{N}=1-\tau_{N}$ is the reflection probability at the normal terminal, $f_0$ is the Fermi function, and $\epsilon$ is the energy. The equation system can be solved analytically, and yields the normalized spectral conductance
\begin{equation}
g(\epsilon) = \frac{\tau^{*}(A^2(\epsilon)r_{N} - (R(\epsilon)-1)(R(\epsilon)r_{N}-1)- A(\epsilon)\tau_{N})}{(A^2(\epsilon)r^2_{N} - (R(\epsilon)r_{N}-1)^2)\tau_{S}} \nonumber
\end{equation}
where $\tau^{*}=\tau_{S}+\tau_{N}-\tau_{S}\tau_{N}$. The superconducting interface is modeled with a generalized BTK model \cite{Blonder1982,Golubov1995,Perez-Willard2004}, which allows us to incorporate the effect of pair-breaking in an applied magnetic field as an additional test of the applicability of the model.

\subsection{Experimental results}

Fig.~\ref{fig:superconducting}(a) shows differential conductance $dI/dV$ curves measured as a function of applied bias voltage $V_{\mathrm{bias}}$ at zero magnetic field and $T \sim$~20~mK under three different gate conditions, \textit{i.e.}\,at $V_\mathrm{g} =9.8$, -9.8, and -2\,V corresponding to the Fermi level sitting in the conduction band, the valence band and close to the charge neutrality point, respectively. In the subgap regime, we observe Andreev reflections giving rise to non-zero conductance. Contrary to the previous reports on NGS junctions, we do not observe any zero-bias anomaly \cite{Han2018,Popinciuc2012} which is usually interpreted as the effect of reflectionless tunelling at the superconductor interface \cite{Kastalsky1991,vanWees1992,Marmorkos1993,Schechter2001}. By using the modified OTBK model to fit the data, we obtain variation of $\Delta$ from $\sim$127~$\mu e$V in the n-doped region to $\sim$144~$\mu e$V in the p-doped region. This difference in $\Delta$ can be seen as the shift in the $dI/dV$ maxima in Fig.~\ref{fig:superconducting}(a) and it can be attributed to two effects. First, the effective temperature of the device is higher on the electron side due to higher normal state conductance as can be seen in Fig.~\ref{fig:Resistance}. It can cause self-heating of the device leading to a decrease in $\Delta$. Second, there could be a gate dependent voltage division taking place between the normal metal lead and the graphene sheet.

\begin{figure}
\includegraphics[width=1\textwidth]{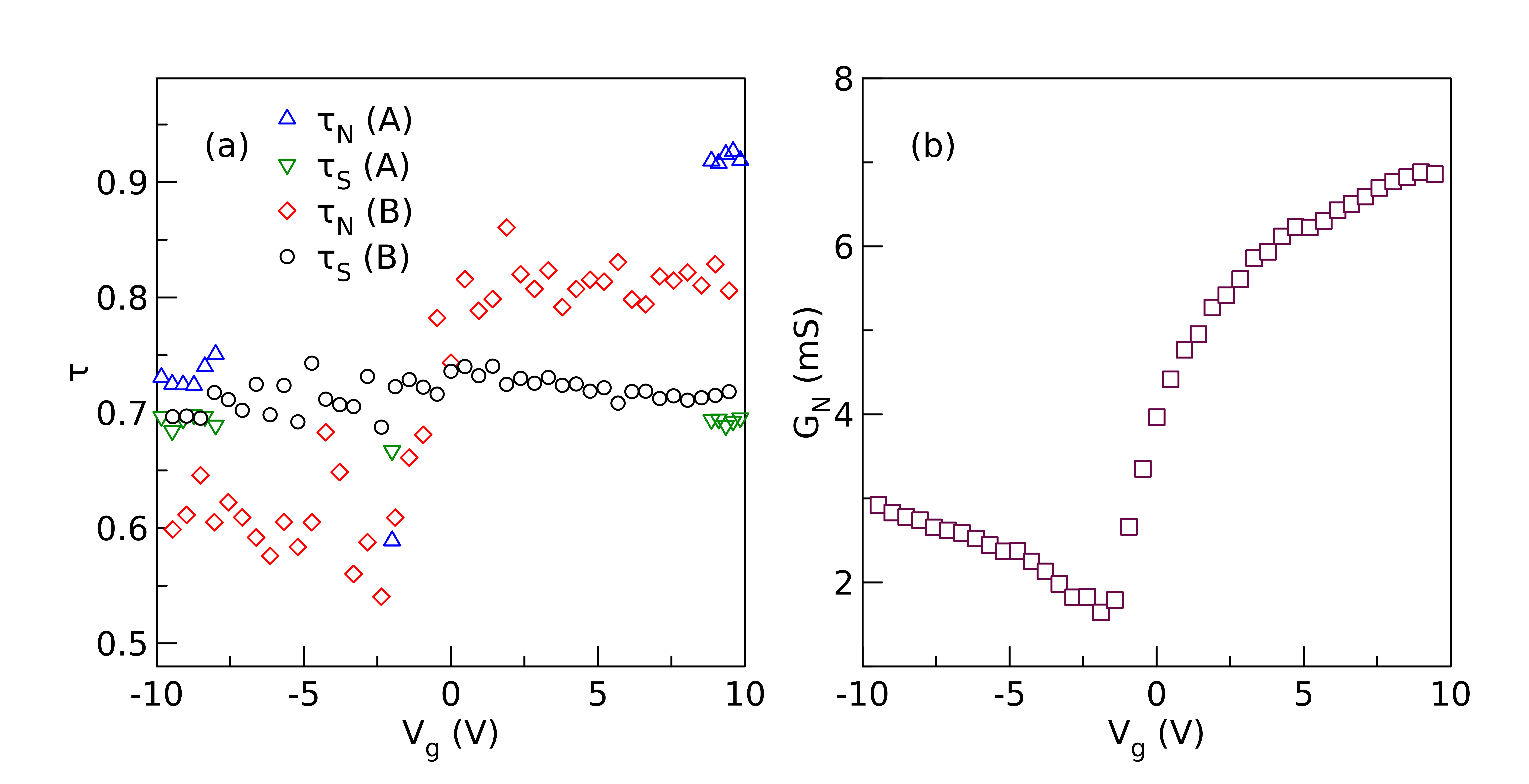}
\caption{(Color online) (a) Transmission probabilities $\tau_\mathrm{N}$ and $\tau_\mathrm{S}$ of the normal metal/graphene and graphene/superconducting interface, respectively for device A and B. (b) Normal state conductance $G_\mathrm{N}$ as a function of gate voltage $V_\mathrm{g}$ covering a large scale of charge carrier density in device B. All the $\tau_\mathrm{N}$, $\tau_\mathrm{S}$ and $G_\mathrm{N}$ are obtained from the best fitting values of differential conductance data measured at a base temperature of 20~mK and $B = 0$~T using our modified OTBK model.}
\label{fig:Transmission}
\end{figure}

In order to test the applicability of our model with respect to an applied magnetic field, we have studied the differential conductance $dI/dV$ as a function of perpendicular magnetic field $B$. Fig.~\ref{fig:superconducting}(b) shows a series of $dI/dV$ curves at various $B$ at $V_\mathrm{g} = -9.2$~V and a base temperature of $T \sim$~20~mK. While the applied magnetic field is kept below the critical magnetic field $B_\mathrm{C}$, thereby preserving superconductivity in the device, a clear decrease in $\Delta$ can be seen in the successive conductance curves. We observe a decrease in $\Delta$ from $\sim$143~$\mu e$V at 0~mT down to $\sim$105~$\mu e$V at 6~mT while the magnetic depairing energy, accounting for magnetic pair-breaking effects \cite{Maki1964,Meservey1975}, changes from 0 to $\sim$0.13$\Delta$.

\begin{figure}
\includegraphics[width=1\textwidth]{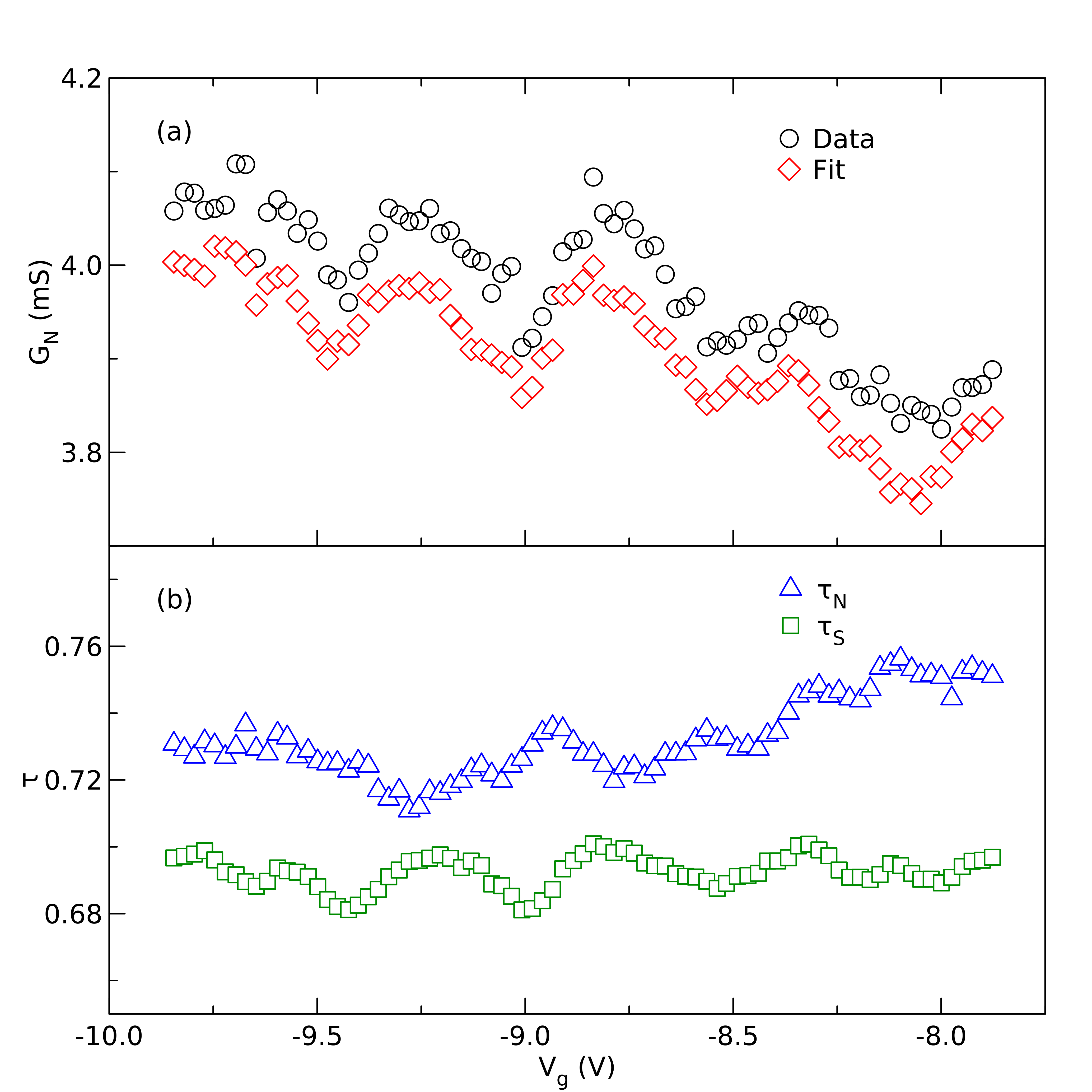}
\caption{(Color online) (a) Oscillations in the normal conductance $G_\mathrm{N}$ observed at a base temperature of 20~mK and $B = 0$~T as a function of gate voltage $V_\mathrm{g}$ in the hole-doped region (circles), and best fitting value for the conductance obtained from the OTBK fits of the experimental data (diamonds). (b) Transmission probability $\tau$ obtained from the OTBK fits of the experimental data.}
\label{fig:correlation}
\end{figure}

Superconductivity weakens as the temperature starts to rise to the critical temperature $T_\mathrm{C}$. In order to see the evolution of superconductivity in our devices, we measured a series of temperature dependent conductance spectra ranging from $T \sim 102$~mK to 780~mK in the hole-doped regime at $V_\mathrm{g} = -9$~V under zero magnetic field. It can be clearly seen in Fig.~\ref{fig:superconducting}(c) that $\Delta$ decreases with increasing temperature and the system goes towards the normal state. For the sake of clarity, the differential conductance curves have been shifted downward by 0.4~mS in successive steps. The series can be well-captured with the fits obtained from the modified OTBK model. With increasing temperature, $\Delta$ decreases from $\sim$144~$\mu e$V to $\sim$119~$\mu e$V. As can be seen, in the low temperature regime, the model provides reasonable fit of the data. However, as the temperature approaches $T_\mathrm{C}$, the data is not very-well fitted by the model. It shows the limitation of the model close to $T_\mathrm{C}$.

In our devices, the two different interfaces, namely the normal metal/graphene and the graphene/superconductor interfaces, play a crucial role in the electronic transport. Fig.~\ref{fig:Transmission}(a) displays the extracted transmission probabilities $\tau_\mathrm{N}$ and $\tau_\mathrm{S}$ of the normal metal/graphene and graphene/superconductor interface, respectively, over a large gate voltage range for device B and in a selected gate voltage range for device A. Both transparencies are extracted from our $dI/dV$ fits. In case of device A, the transmission probability for the normal metal/graphene interface, $\tau_\mathrm{N}$, obtained from the fits varies from 0.74 in the p-doped region to 0.92 in the n-doped region, dropping to 0.59 close to the charge neutrality point. The transmission probability for the graphene/superconductor interface, $\tau_\mathrm{S}$, on the other hand, varies from 0.67 to 0.71 for the entire gate voltage range indicating a weak barrier at the interface. In case of device B, we have obtained lower transmission probability for the normal metal/graphene interface as compared to device A, however, the dependence of $\tau_\mathrm{N}$ and $\tau_\mathrm{S}$ on the applied gate voltage follows the same trend as device A. Note that this systematic dependence of the transmission probabilities on doping is directly visible in the change of the normalized subgap conductance in Fig.~\ref{fig:superconducting}(a). The difference between $\tau_\mathrm{N}$ and $\tau_\mathrm{S}$ extracted with our model confirms the asymmetry of the interference patterns observed in the differential conductance map in the normal state presented in Fig.~\ref{fig:FigFP}(a). Fig.~\ref{fig:Transmission}(b) shows the extracted normal state conductance $G_\mathrm{N}$  for device B as a function of gate voltage $V_\mathrm{g}$ where a large difference between hole and electron conductance is observed due to the n-doping of the contact as mentioned before. 

For a more detailed analysis of the correlation between the gate dependent normal state conductance and transmission probabilities, we measured a series of conductance spectra at $B = 0$~T with varying $V_\mathrm{g}$ in the p-doped region at high charge carrier densities for device A. Fig.~\ref{fig:correlation}(a) shows the normal conductance $G_\mathrm{N}$ as circular data points obtained from these measurements as a function of $V_\mathrm{g}$. The values of $G_\mathrm{N}$ obtained as a fitting parameter from the OTBK fits of the experimental data are plotted as diamonds in Fig.~\ref{fig:correlation}(a) showing oscillations similar to the FP interference observed in Fig.~\ref{fig:Resistance}. In Fig.~\ref{fig:correlation}(b), the transmission probabilities obtained from the same fits are shown. It is clear that our fits can qualitatively describe the observed conductance oscillations in terms of transmission probability. It suggests that the transmittance of the interfaces is tuned by $V_\mathrm{g}$ in a similar manner as observed for the FP interference. The oscillation period observed here also corresponds to a cavity length $L_\mathrm{C}$ of $\sim$ 236 $\pm$ 10 ~nm. It proves that these oscillations indeed arise from the FP interference of the Andreev reflected charge carriers. In a Josephson junction, existence of FP interference can be observed in the superconducting state by following the oscillations in the critical current \cite{Calado2015,BenShalom2016,Amet2016,Borzenets2016,Zhu2016,Allen2017,Nanda2017,Zhu2018,Park2018,Kraft2018,Schmidt2018,Kroll2018,Draelos2019,Wang2019} or multiple Andreev reflection \cite{Allen2017}. However, to the best of our knowledge, it is the first time that the effect of FP resonances have been observed in superconducting state in NGS junctions.

\section{Conclusion}

To conclude, we have reported a study of normal metal/graphene/superconductor junctions in the ballistic regime. We see that in the normal state, the differential conductance shows asymmetric Fabry-P\'{e}rot interference when a cavity is formed by charge transfer at the contacts. We have attributed the asymmetry to the different transmissions  of the two metal/graphene contact interfaces in our devices. In the superconducting state, the junctions are well described by our modified OTBK model. The Fabry-P\'{e}rot interference of the electronic transport is directly reflected in the transmission probabilities in these systems.

\acknowledgements{The authors thank R. M\'{e}lin and C. Sch\"{o}nenberger for fruitful discussions. This work was partly supported by Helmholtz society through program STN and the DFG via the projects DA 1280/3-1. P.P. acknowledges support from Deutscher Akademischer Austauschdienst (DAAD) scholarship.}

\end{document}